\documentclass[fleqn,11pt]{elsarticle}
\usepackage{lineno,graphicx,caption,hyperref}
\hypersetup{
colorlinks=true,
 citecolor=blue,
 linkcolor=blue,
 urlcolor=blue}
\modulolinenumbers[1]
\journal{arXiv}
\usepackage{fullpage}
\usepackage{xparse, amsmath, amssymb, physics, xcolor}
\biboptions{sort&compress}

\newcommand{\equa}[1]{equation~(\ref{#1})}

\newcommand{\equas}[1]{equations~(\ref{#1})}
\newcommand{\equass}[2]{equations~(\ref{#1})--(\ref{#2})}
\newcommand{\equasa}[2]{equations~(\ref{#1}){ }and{ }(\ref{#2})}

\newcommand{\Equas}[1]{Equations~(\ref{#1})}

\newcommand{\eqn}[2]{\begin{gather}
\begin{aligned}
#1
\label{#2}
\end{aligned}
\end{gather}
}
\newcommand{\gat}[2]{\begin{subequations}\label{#2}\begin{gather}
#1
\end{gather}\end{subequations}
}

\newcommand{\spl}[2]{\begin{multline}
#1
\hfill
\label{#2}
\end{multline}
}

\begin{document}

\begin{frontmatter}

\title{{\bf Thermal instability of the buoyant flow in a vertical cylindrical porous layer with a uniform internal heat source}}

\author{A.\ Barletta}
\corref{mycorrespondingauthor}
\cortext[mycorrespondingauthor]{Corresponding author
}
\ead{antonio.barletta@unibo.it}

\address{Department of Industrial Engineering, Alma Mater Studiorum Universit\`a di Bologna,\\ Viale Risorgimento 2, 40136 Bologna, Italy}

\author{D.A.S.\ Rees}
\ead{ensdasr@bath.ac.uk}

\address{Department of Mechanical Engineering, University of Bath,\\ Claverton Down, Bath BA2 7AY, United Kingdom}

\author{B.\ Pulvirenti}
\ead{beatrice.pulvirenti@unibo.it}

\address{Department of Industrial Engineering, Alma Mater Studiorum Universit\`a di Bologna,\\ Viale Risorgimento 2, 40136 Bologna, Italy}

\begin{abstract}
The buoyancy--induced parallel flow in a vertical cylindrical porous layer is analysed. A radial thermal gradient caused by a uniformly distributed heat source is assumed to induce the buoyant flow. The layer boundaries are modelled as isothermal and permeable to an external fluid reservoir. The onset of the convective instability is analysed by linearising the governing equations for the perturbations. The governing parameters driving the instability are the heat--source Rayleigh number and the ratio between the internal radius and the external radius. Neutral stability curves and the critical values of the Rayleigh number, the perturbation wave number and the angular frequency are computed numerically. It is shown that axisymmetric modes form the most dangerous mode of instability.
\end{abstract}


\end{frontmatter}

\section{Introduction}
The onset of thermal instability in a fluid--saturated porous medium has a widespread interest in the heat transfer community, as may be inferred from the quite abundant literature on this topic; see \citet{NieldBejan2017}. There are several areas of engineering and physics within which such investigations are applied. We just mention the analysis of contaminant diffusion in the soil, the extraction of hydrocarbons, the $\rm CO_2$ sequestration processes, and the use of metal foams for the optimised design of heat exchangers. If most of the studies published in the last decades are focussed on the thermal instability of the Rayleigh--B\'enard type, where the fluid is initially at rest while experiencing a purely conductive heat transfer, there are other analyses devoted to side heating conditions or internal heating conditions in vertical porous layers, where a stationary and parallel buoyant flow may give rise to a multi--cellular instability pattern. A survey of the latter type of instability may be found in Chapter~7 of \citet{NieldBejan2017}. Recent important results have been discussed by several authors \citep{Rees11, scott2013nonlinear, Barletta2015, shankar2017stability, barletta2017instability, naveen2020finite, shankar2020impact}. In particular, \citet{barletta2017instability} proved that the parallel buoyant flow in a plane vertical porous layer with a uniform internal heat source may become unstable even in the absence of a temperature difference between the boundaries.  

The aim of this paper is to develop and extend the stability analysis presented in \citet{barletta2017instability} by investigating the effect of curvature when a vertical annular porous layer is considered instead of a plane vertical layer. The plane layer behaviour of \citet{barletta2017instability} is then found as a limiting case where the aspect ratio between the internal radius and the external radius of the annulus tends to unity. The present stability analysis is carried out by assuming small--amplitude perturbations of the basic buoyant flow. The linear dynamics of perturbations is determined by employing a modal analysis. The resulting eigenvalue problem is solved numerically, thus providing the neutral stability curves and the critical values for the onset of the instability as a function of the aspect ratio of the annulus.

\section{Mathematical model}
In analogy with the system studied by \citet{barletta2017instability}, we consider a vertical porous annulus with infinite height, internal radius $r_1$ and external radius $r_2$. A fluid saturates the porous medium. Cylindrical coordinates $(r,\phi,z)$ are chosen so that the vertical coordinate, $z$, is also the axis of the cylinder. A uniform internal heat source, with power per unit volume $\dot{q}$, is present inside the annulus. We can devise conditions such that $\dot{q}$ is caused by the Joule heating due to a stationary electric current in the porous medium or, alternatively, caused by an exothermic chemical reaction. The boundaries $r=r_1$ and $r=r_2$ are considered to be both isothermal and isobaric, with the temperature, $T_s$, and the pressure equal to the hydrostatic pressure of the fluid. The latter condition models perfect permeability of the boundary to an external fluid reservoir at rest. By introducing the local difference between the pressure and the hydrostatic pressure, $p$, the boundary conditions are that $p=0$ at  both $r=r_1$ and $r=r_2$.

\subsection{Governing equations}
The governing equations for the seepage flow in the porous cylinder are based on the Oberbeck--Boussinesq approximation and on Darcy's law \citep{NieldBejan2017}. Hence, we write
\gat{
\div{\vb{u}} = 0 , \label{1an}\\
\frac{\mu}{K} \, \vb{u} = -\, \grad{p} + \rho g \beta \qty(T - \bar{T}) \vu{e}_z, \label{1bn}\\
\sigma \pdv{T}{t} + \vb{u} \vdot \grad{T} = \alpha\qty( \nabla^2 T + \frac{\dot{q}}{\lambda}) , \label{1cn}
}{1n}
which express
the local mass balance \equa{1an}, the local momentum balance \equa{1bn} and the heat transport \equa{1cn}. In \equas{1n}, $\vb{u}$ is the seepage velocity with components $(u,v,w)$ along the $(r,\phi,z)$ directions, $T$ is the temperature field and $t$ is the time. Furthermore, in \equas{1n}, $\mu$, $\beta$ and $\rho$ are the fluid dynamic viscosity, thermal expansion coefficient and reference density, while $\alpha$ is the average thermal diffusivity of the saturated medium, $\lambda$ is the average thermal conductivity of the saturated porous medium, $K$ its permeability and $\sigma$ the ratio between the volumetric heat capacity of the saturated porous medium and that of the fluid. The modulus of the gravitational acceleration is $g$ and $\vu{e}_z$ is the unit vector along the $z$ axis.
The constant $\bar{T}$ denotes the average temperature in an annular cross--section $(z = constant)$ evaluated for the basic state to be defined in the forthcoming Section~\ref{basbuoflo}.
The local balance \equas{1n} can be rewritten in a dimensionless form by means of the scaling
\spl{
\frac{1}{r_2}\ (r, z) \to (r, z) \qc \frac{\alpha}{\sigma r_2^2}\ t \to t \qc \frac{K}{\mu \alpha}\ p \to p,\\
\frac{r_2}{\alpha}\ \vb{u} = \frac{r_2}{\alpha}\ (u, v, w) \to (u, v, w) = \vb{u} \qc  
\lambda\,\frac{T - \bar{T}}{\dot{q} r_2^2} \to T.
}{2}
The Rayleigh number $R$ is defined as
\eqn{
R = \frac{\rho g \beta \dot{q} K r_2^3}{\lambda \mu \alpha},
}{3}
using the outer radius, $r_2$, as the length scale. From \equass{1n}{3}, we obtain
\gat{
\div{\vb{u}} = 0 , \label{1a}\\
\vb{u} = -\, \grad{p} + R\, T \, \vu{e}_z, \label{1b}\\
\pdv{T}{t} + \vb{u} \vdot \grad{T} = \nabla^2 T + 1 . \label{1c}
}{1}
We mention that the dimensional average temperature $\bar{T}$ defines the reference temperature within the Oberbeck--Boussinesq approximation. Thus, when either $T=\bar{T}$ (by using dimensional temperatures) or $T=0$ (by using the dimensionless temperature), the buoyancy force is zero.

\subsection{Boundary conditions}
In dimensionless form, the pressure and temperature boundary conditions are expressed as
\eqn{
p = 0 \qc T = a \qq{at} r = \gamma \qq{and} r = 1,
}{5}
where $\gamma$ is the aspect ratio and $a$ is a dimensionless parameter given by
\eqn{
\gamma = \frac{r_1}{r_2} \qc a = \lambda\,\frac{T_s - \bar{T}}{{\dot{q} r_2^2}}.
}{6}
As it will become clearer in the next Section~\ref{basbuoflo}, the value of $a$ depends on the net flow rate across the porous annulus.

\subsection{Basic buoyant flow}\label{basbuoflo}
A steady parallel flow in the vertical $z$ direction exists. It is defined by the solution of \equasa{1}{5} and expressed as
\spl{
u_b = 0 = v_b \qc w_b(r) = R\, \qty[a + F(r)] ,
\\
T_b(r) = a + F(r) \qc p_b = 0,
\qq{with} F(r) = \frac{\qty(1 - r^2) \ln\!\qty(\gamma) - \qty(1 - \gamma ^2) \ln\!\qty(r)}{4 \ln\!\qty(\gamma)} .
}{7}
Here, the subscript ``$b$'' serves to denote the ``basic'' flow. The flow is caused entirely by the buoyancy force as may easily be inferred from the velocity being proportional to the Rayleigh number $R$. The value of the parameter $a$ is correlated to the flow rate across a $z=constant$ cross--section,
\eqn{
\bar{w}_b = \frac{2}{1 - \gamma^2} \int_{\gamma}^1 w_b\, r\, \dd r = R\; \qty[a + \frac{1 - \gamma^2 + \qty(1 + \gamma^2) \ln\!\qty(\gamma)}{8 \ln\!\qty(\gamma)}] .
}{8}
There exists a special case, 
\eqn{
a = - \frac{1 - \gamma^2 + \qty(1 + \gamma^2) \ln\!\qty(\gamma)}{8 \ln\!\qty(\gamma)} ,
}{9}
which defines a condition of zero flow rate in the basic state. Such a condition corresponds to when the fluid is confined within an annulus which is very considerably taller than its outer radius. We point out that \equasa{6}{9} implicitly define the constant reference temperature $\bar{T}$ employed in \equa{1bn}.

\subsection{Pressure--temperature formulation}
By evaluating the divergence of \equa{1b} and by employing \equa{1a}, we can rewrite \equasa{1}{5} as
\gat{
\nabla^2 p = R\; \pdv{T}{z} , \label{9a}\\
\pdv{T}{t} - \grad{p} \vdot \grad{T} + R\, T\; \pdv{T}{z} = \nabla^2 T + 1 , \label{9b}\\
p = 0 \qc T = a \qq{at} r = \gamma, 1. \label{9c}
}{10}
The advantage in the formulation (\ref{10}) relies on the reduced number of unknowns $(p,T)$ to be determined with respect to \equas{1}, where the unknowns are $(\vb{u}, p, T)$.

\section{Linear stability analysis}
It is well-known that stationary solutions of the governing equations might be unstable under certain parametric conditions. In our case, the basic flow (\ref{7}) may be stable or unstable depending on the parameters $\gamma$ and $R$. By introducing the perturbation parameter $\varepsilon$, where $|\varepsilon| \ll 1$ is assumed, we will carry out a linear stability analysis of the perturbations superposed onto the basic flow (\ref{7}). Hence, by employing \equa{7}, we write
\eqn{
p(r,\phi,z,t) = \varepsilon P(r,\phi,z,t) \qc T(r,\phi,z,t) = a + F(r) + \varepsilon \theta(r,\phi,z,t) ,
}{11}
where $\qty(P,\theta)$ are the perturbations. If we substitute \equa{11} into \equas{10} and if we neglect terms of $O\qty(\varepsilon^2)$, then we obtain the linearised governing equations for the unknowns $\qty(P,\theta)$, namely
\gat{
\nabla^2 P = R\; \pdv{\theta}{z} , \label{12a}\\
\pdv{\theta}{t} - F'(r)\; \pdv{P}{r} + R\, \qty[a + F(r)]\; \pdv{\theta}{z} = \nabla^2 \theta , \label{12b}\\
P = 0 \qc \theta = 0 \qq{at} r = \gamma, 1, \label{12c}
}{12}
where primes serve to denote derivatives with respect to $r$.
The dependence on the angular coordinate $\phi$ can be managed by using the Fourier series,
\spl{
P\qty(r,\phi,z,t) = \sum_{n=0}^\infty P_n\qty(r,z,t) \cos\!\qty(n\phi) ,
\\
\theta\qty(r,\phi,z,t) = \sum_{n=0}^\infty \theta_n\qty(r,z,t) \cos\!\qty(n\phi) .
}{13}
Thus, we obtain for $n=0,1,2,\ \ldots$
\gat{
\frac{1}{r}\, \pdv{r} \qty( r\, \pdv{P_n}{r}) + \pdv[2]{P_n}{z} - \frac{n^2}{r^2} P_n = R\, \pdv{\theta_n}{z} , \label{14a}\\
\frac{1}{r}\, \pdv{}{r} \qty( r\, \pdv{\theta_n}{r}) + \pdv[2]{\theta_n}{z} - \frac{n^2}{r^2} \theta_n = \pdv{\theta_n}{t} - F'(r)\; \pdv{P_n}{r} + R\, \qty[ a + F(r)]\; \pdv{\theta_n}{z} , \label{14b}\\
P_n = 0 \qc \theta_n = 0 \qq{at} r = \gamma, 1. \label{14c}
}{14}
We now focus on the dynamics of normal modes expressed as
\eqn{
P_n\qty(r,z,t) = f_n(r) \; e^{\eta t}\, e^{i k z} \qc \theta_n\qty(r,z,t) = h_n(r) \; e^{\eta t}\, e^{i k z} ,
}{15}
with a real wave number, $k$, and the complex growth rate, $\eta$. The substitution of \equa{15} into \equa{14} yields
\gat{
f''_n + \frac{1}{r}\, f'_n - \qty(\frac{n^2}{r^2} + k^2)\, f_n - i k R\, h_n = 0 , \label{16a}\\
h''_n + \frac{1}{r}\, h'_n - \qty[\frac{n^2}{r^2} + k^2 + H + i k R \, F(r)]\, h_n + F'(r)\, f'_n = 0 , \label{16b}\\
f_n = 0 \qc h_n = 0 \qq{at} r = \gamma, 1 , \label{16c}
}{16}
where we defined the modified complex parameter $H$ as
\eqn{
H = \eta + i a k R  .
}{17}
The real part of $H$ coincides with the real part of $\eta$ and, hence, with the growth rate of the normal mode. If we denote with $\xi = \Re(H) = \Re(\eta)$ the exponential growth rate, then $\xi > 0$ defines instability, $\xi < 0$ stability and $\xi = 0$ neutral stability. The imaginary part of $\eta$ is equal to $-\omega$, where $\omega$ is the angular frequency of the normal mode. We can denote the imaginary part of $H$ as $-\Omega$. Thus, by employing \equa{17}, we can write
\eqn{
\Omega = \omega - a k R  .
}{18}
\Equas{16} form a system of homogeneous ordinary differential equations with homogeneous boundary conditions. In fact, \equas{16} yield an eigenvalue problem where the eigenfunctions $(f_n, h_n)$ are to be numerically computed together with the complex eigenvalue $H$, for every prescribed input parameters $(n, k, R)$. As a consequence of the definition (\ref{17}), the parameter $a$ is not involved explicitly in the solution of the eigenvalue problem. In particular, this means that the eigenvalue $H$ is independent of $a$.

The numerical solution of \equas{16} as a differential eigenvalue problem is performed via the shooting method using an adaptive grid. 
We omit here the details of this procedure which is described in Chapter~9 of the book by \citet{Straughan} and in Chapter~10 of the book by \citet{barletta2019routes} where details about the coding of the numerical solver are also provided.
The minimum of the neutral curve is obtained by means of an extended system as described in  \citet{barletta2019routes}.
Numerical data quoted below are accurate to six significant figures.

\begin{table}[t]
\centering
\begin{tabular}{l | l | l l l l l l}
$\gamma$ & $n$ & $\hat{R}_c$ & $\hat{k}_c$ & $\hat{\Omega}_c$\\
\hline
0.99 & 0 & 740.026 & 1.94671 & 197.192\\
     & 1 & 740.035 & 1.94671 & 197.194\\
     & 2 & 740.065 & 1.94670 & 197.201\\
     & 3 & 740.114 & 1.94669 & 197.211\\
\hline
0.75 & 0 & 738.909 & 1.94545 & 197.228\\
     & 1 & 747.021 & 1.94396 & 198.914\\
     & 2 & 771.981 & 1.93864 & 204.022\\
     & 3 & 815.831 & 1.92691 & 212.704\\
\hline
0.5  & 0 & 734.050 & 1.93913 & 197.448\\
     & 1 & 782.121 & 1.92855 & 207.350\\
     & 2 & 952.123 & 1.86815 & 238.931\\     
     & 3 & 1388.49 & 1.65609 & 300.100\\
\hline
0.25 & 0 & 722.518 & 1.91299 & 198.827\\
     & 1 & 935.255 & 1.84584 & 240.859\\
     & 2 & 3212.34 & 1.03842 & 440.713\\
\end{tabular}
\caption{\label{tab1}Critical values of $\hat{R}$, $\hat{k}$ and $\hat{\Omega}$, for some values of $\gamma$ and $n$.}
\end{table}

\section{Discussion of the results}
The onset of the convective instability is identified by the neutral stability curve, defined as the locus in the $(k,R)$ plane where $\xi = \Re(H) = 0$. However, since the comparison with the instability observed in the case of a vertical plane layer is important, it is quite convenient to rescale the pertinent parameters governing the transition to instability,
\eqn{
\hat{R} = \qty(1 - \gamma)^3 R \qc \hat{k} = \qty(1 - \gamma) k \qc \hat\Omega = \qty(1 - \gamma)^2 \Omega  .
}{19}
Such a rescaling is motivated by the change of the reference length from $r_2$, employed in \equa{2}, to the thickness $r_2 - r_1$, which is the equivalent of the natural reference length for a plane layer \citep{barletta2017instability}. By adopting the rescaled parameters defined by \equa{19}, the results for the plane layer are retrieved through the asymptotic solution for $\gamma \to 1$. We recall that, according to \citet{barletta2017instability}, the asymptotic case $\gamma \to 1$ features
\eqn{
\hat{R}_c = 740.027 \qc \hat{k}_c = 1.94671 \qc \hat\Omega_c = 197.192  ,
}{20}
where the subscript ``$c$'' denotes the ``critical'' condition, namely the minimum $R$ position along the neutral stability curve drawn in the $(k,R)$ plane, for a given $\gamma$.

\begin{figure}[t]
\centering
\includegraphics[width=0.9\textwidth]{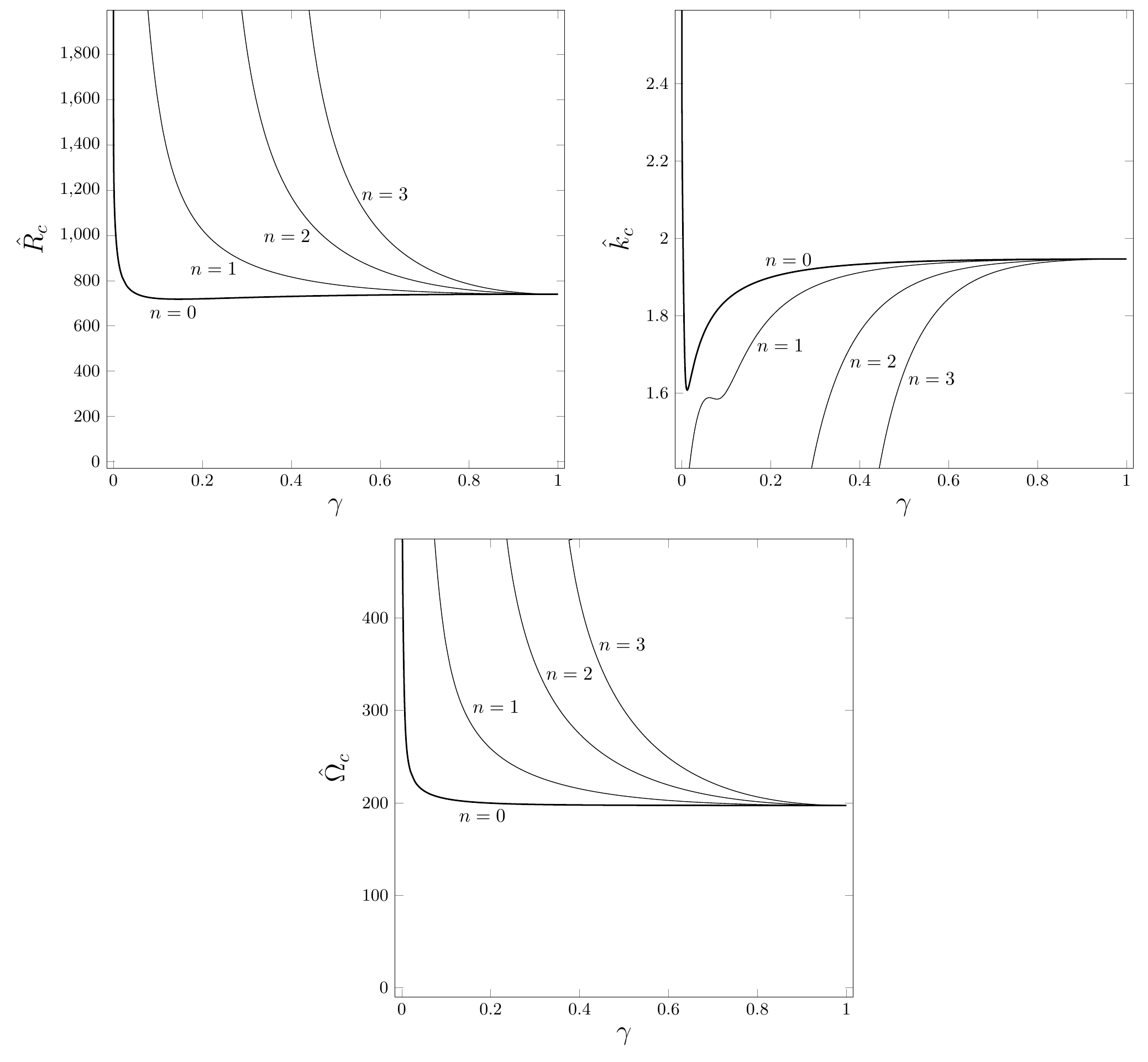}
\captionsetup{aboveskip=0pt}
\caption{\label{fig1}Variation of $\hat{R}_c$, $\hat{k}_c$ and $\hat{\Omega}_c$ wth $\gamma$ for $n=0$ (thick line) and for $n=1,2,3$ (thin lines).}
\end{figure}

Table~\ref{tab1} reports some values of $(\hat{R}_c, \hat{k}_c, \hat{\Omega}_c)$ versus $\gamma$ and $n$. The general evidence is that the dependence on $n$ is very weak when the annulus has a small curvature $(\gamma=0.99)$ even if, also in this case, $\hat{R}_c$ increases with $n$. On the other hand, the dependence on $n$ becomes more and more dramatic as $\gamma$ decreases. This phenomenon is apparent especially with regard to the values of $\hat{R}_c$. In the case $\gamma=0.25$, reported in Table~\ref{tab1}, the critical values for $n=3$ could not be computed and, hence, they are omitted in the table. A possible reason is that, with $\gamma=0.25$ and $n=3$, the value of $\hat{R}_c$ becomes so large that numerical accuracy is lost. Moreover, Table~\ref{tab1} shows that the critical values for $\gamma=0.99$ and $n=0$ coincide to within six significant figures with the asymptotic values obtained by \citet{barletta2017instability} and reported above in \equa{20}. Table~\ref{tab1} suggests that the smallest value of $\hat{R}_c$ corresponds to when $n=0$ (axisymmetric modes) and that it decreases with $\gamma$. 
Thus, a departure from the plane layer geometry by having an increased curvature of the annulus causes the basic flow to be destabilised at decreasing values of $\hat{R}_c$.

\begin{figure}[t]
\centering
\includegraphics[width=0.9\textwidth]{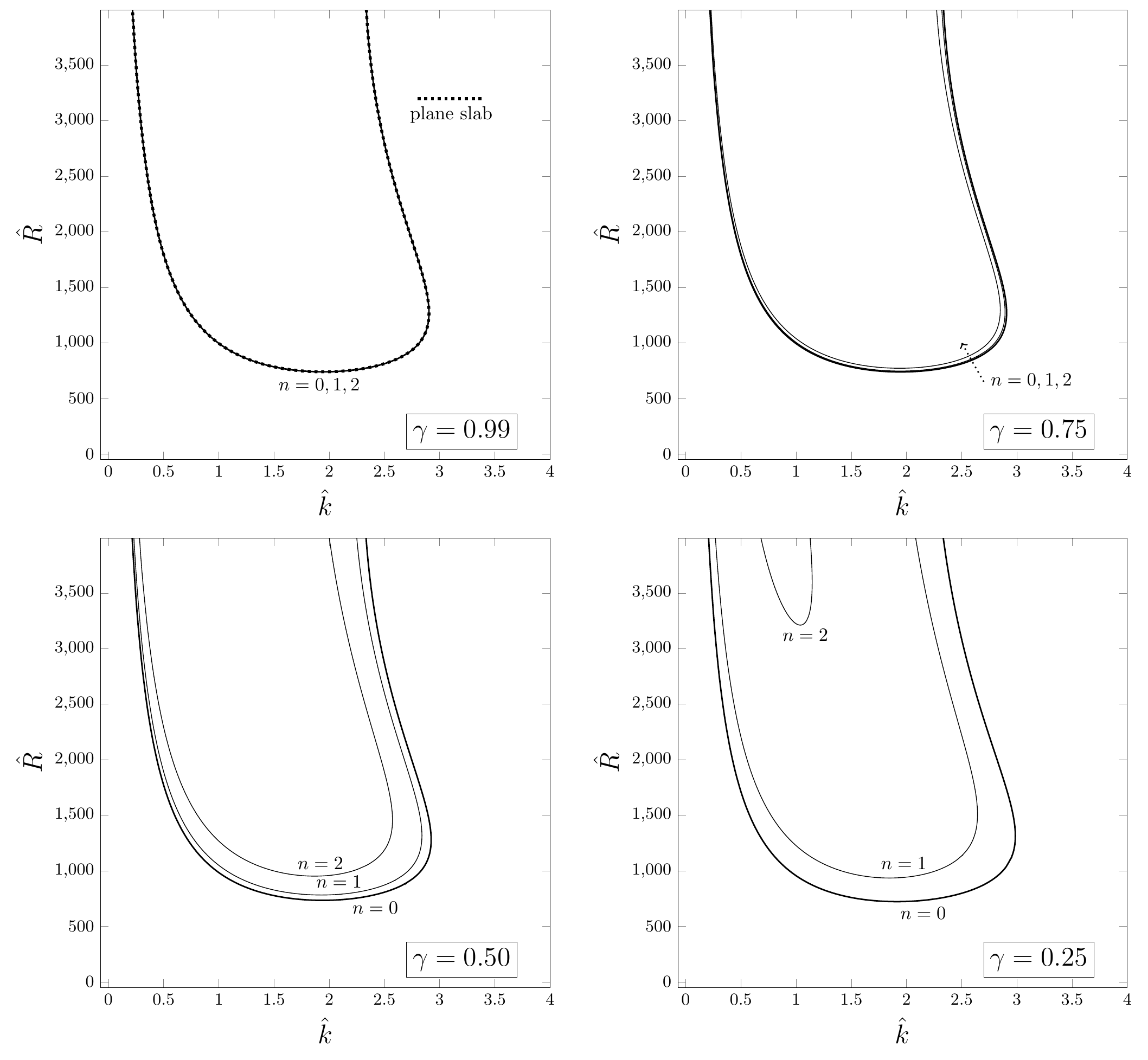}
\captionsetup{aboveskip=0pt}
\caption{\label{fig2}Neutral stability curves in the $(\hat{k}, \hat{R})$ plane for $n=0$ (thick line) and for $n=1,2$ (thin lines).}
\caption*{Different frames correspond to different aspect ratios, $\gamma$. The dotted line in the frame} 
\caption*{with $\gamma=0.99$ corresponds to the data for the plane slab \citep{barletta2017instability}.}
\end{figure}

Figure~\ref{fig1} illustrates how $\hat{R}_c$ varies with $\gamma$ for $n=0,1,2,3$. As has already been commented on when discussing the data in Table~\ref{tab1}, the modes with $n=0$ are those which trigger the instability at the smallest value of $\hat{R}_c$. The value of $\hat{R}_c$ for $n=0$ decreases slightly when $\gamma$ decreases from $1$ to $\gamma=0.146052$, and thereafter $\hat{R}_c$ increases rapidly  as $\gamma$ decreases still further. The minimum value which occurs at $\gamma=0.146052$ is $\hat{R}_c = 718.208$. Generally, Figure~\ref{fig1} shows that $\hat{k}_c$ decreases and $\hat{\Omega}_c$ increases as $\gamma$ decreases from $1$, although 
there is an exception in a narrow region with $0<\gamma<0.011879$ where $\hat{k}_c$ increases once more as $\gamma$ decreases.
We reckon that, with such small values of $\gamma$, the sensitivity to the change of the aspect ratio is mainly due to the boundary conditions at the inner boundary which may turn out to be poorly realistic in the limit $\gamma \to 0$.

\begin{figure}[ht!]
\centering
\includegraphics[width=0.95\textwidth]{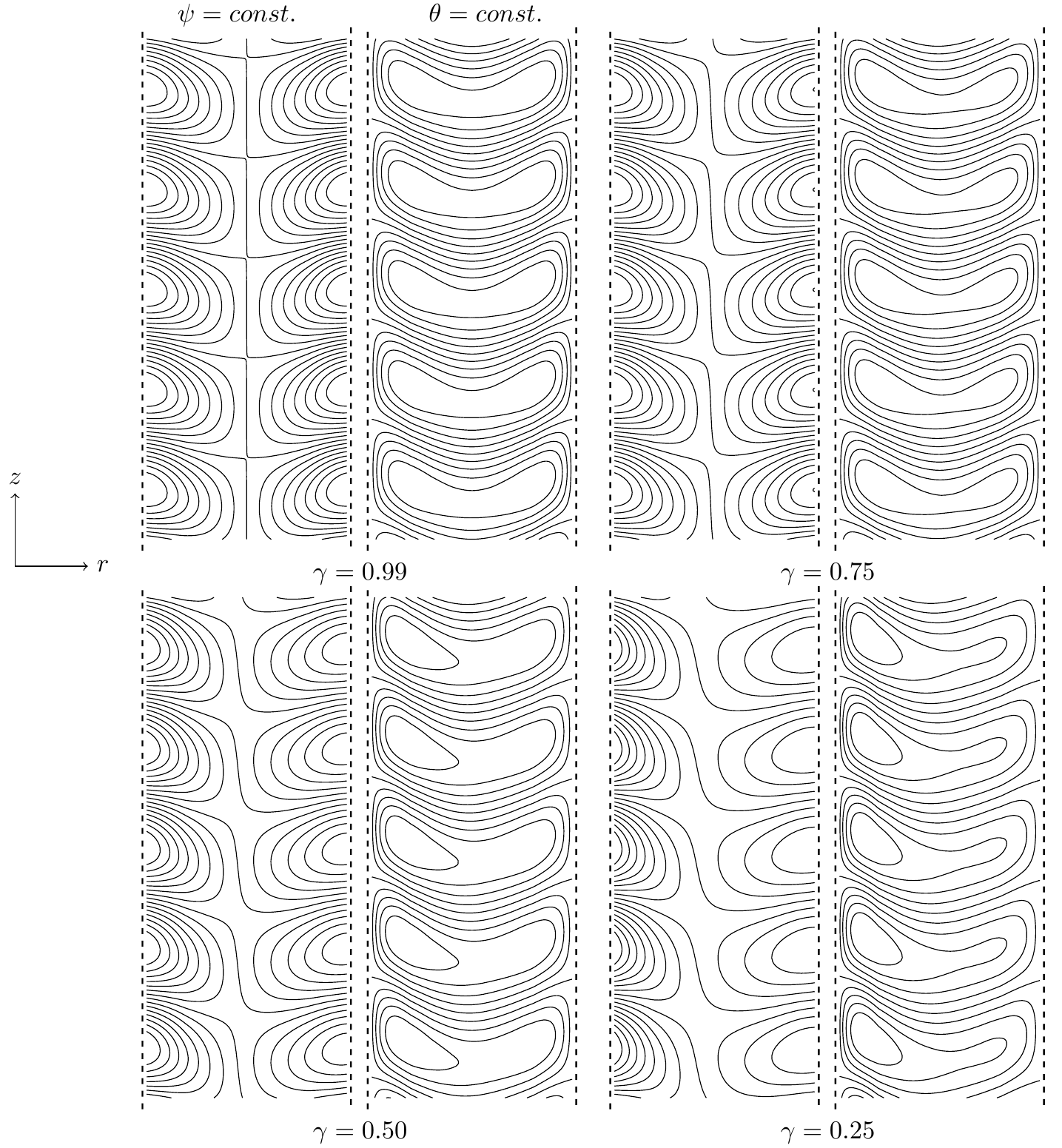}\\[6pt]
\captionsetup{aboveskip=0pt}
\caption{\label{fig3}Streamlines $(\psi=constant)$ and isotherms $(\theta=constant)$ in the $(r,z)$ plane for the axisymmetric}
\caption*{perturbation modes $(n=0)$ with critical conditions and different aspect ratios $\gamma$.}
\end{figure}

The critical values discussed so far result from minimising $\hat{R}$ in the $(\hat{k},\hat{R})$ plane along the neutral stability curves. Then, Figure~\ref{fig2} displays the neutral stability curves in the $(\hat{k},\hat{R})$ plane for a few sample aspect ratios, $\gamma=0.99, 0.75, 0.5, 0.25$, illustrating the effect of a gradual departure from the zero curvature limit, $\gamma \to 1$, analysed by \citet{barletta2017instability}. The behaviour is monitored for the modes $n=0,1,2$ as higher values of $n$ yield higher threshold values of $\hat{R}$ for the transition to instability, as already illustrated through Table~\ref{tab1} and Figure~\ref{fig1}. As evidenced above, the effect of $n$ is extremely small, hardly visible, when $\gamma$ is close to $1$, while this effect is more and more significant as $\gamma$ decreases. In the frame for $\gamma=0.99$, the curves with $n=0,1,2$ cannot be distinguished. On the other hand, they are substantially different from one another when $\gamma=0.25$. For this value of $\gamma$, there is a very large difference between the neutral stability curves with $n=1$ and $n=2$. This result indicates that, when $\gamma$ is small, non--axisymmetric modes act in a markedly different way with respect to each other and to the axisymmetric modes, which is not the case for $\gamma \to 1$. Finally, we mention that Figure~\ref{fig2} shows clearly that the data for $\gamma=0.99$ yield neutral stability curves which are almost indistinguishable from one another, and from the one (dotted curve) reported in \citet{barletta2017instability} for the plane slab case, {\em i.e.} the limit $\gamma\to 1$.

\begin{figure}[t!]
\centering
\includegraphics[width=0.95\textwidth]{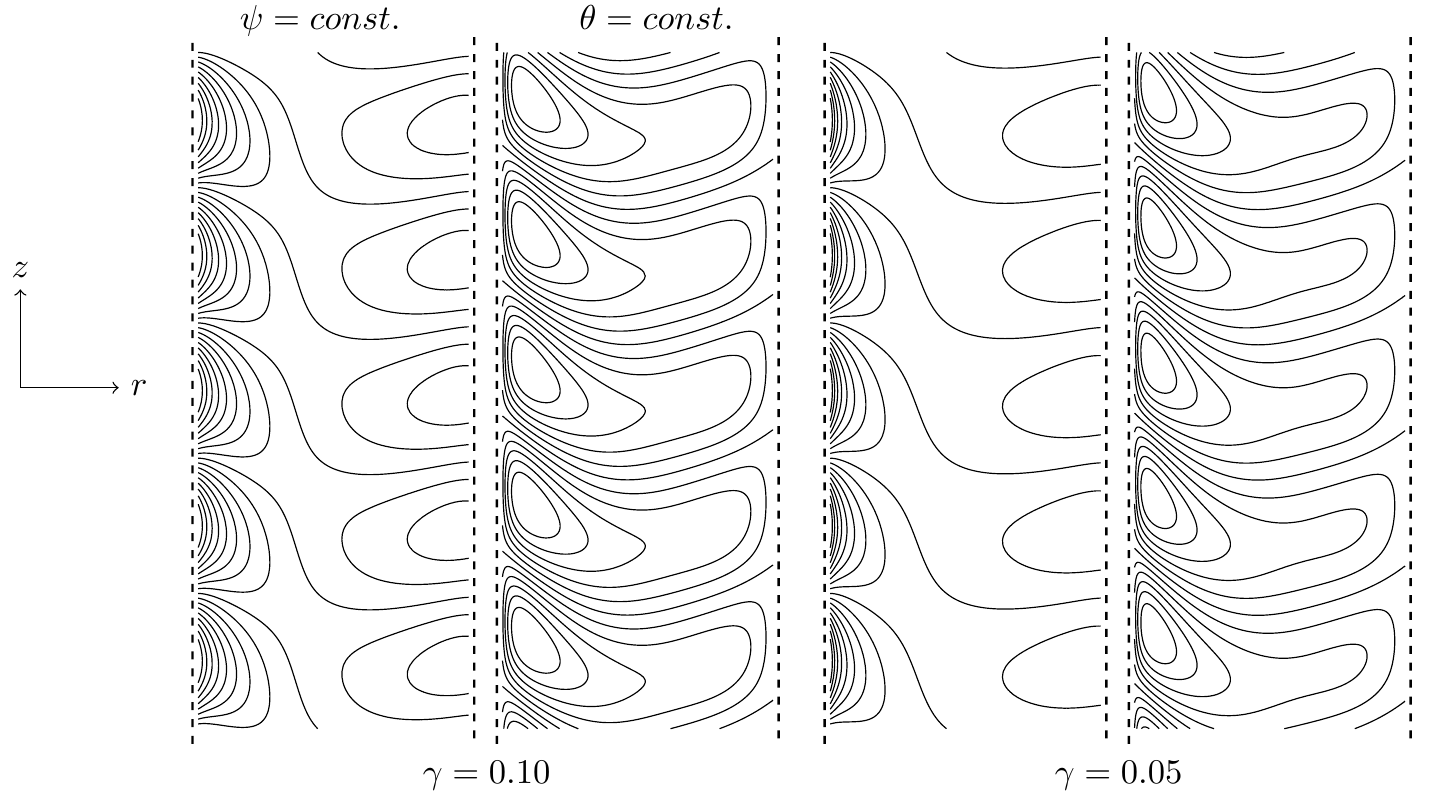}\\[6pt]
\captionsetup{aboveskip=0pt}
\caption{\label{fig4}Streamlines $(\psi=constant)$ and isotherms $(\theta=constant)$ in the $(r,z)$ plane for the axisymmetric}
\caption*{perturbation modes $(n=0)$ with critical conditions and different aspect ratios $\gamma$.}
\end{figure}

{A physical argument justifying the apparent equivalence of different $n>0$ modes at onset of instability when $\gamma$ is very close to unity is as follows.
The dimensionless arc length of a cell in the $(r,\phi)$ plane, $L_{\phi}$, can be evaluated by employing \equa{13}. It depends on the nonzero value of $n$ and is given by, approximately,
\eqn{
L_{\phi} = \frac{\pi \qty(1 + \gamma)}{2 n},
}{21}
where we have assumed the reference radius as the arithmetic mean between the external and internal radii.
$L_{\phi}$ is to be compared with the radial width of the cell,
\eqn{
L_r = 1 - \gamma.
}{22}
If $n$ is such that $L_r \ll L_{\phi}$, then the non-axisymmetric modes are
comparable with the axisymmetric modes in the onset of instability. Thus, in order to see a significantly strong effect of $n$, one should consider modes with
\eqn{
n \sim \frac{\pi \qty(1 + \gamma)}{2 \qty(1 - \gamma)},
}{23}
or larger. Such $n$ is greater than $10^2$ when $\gamma=0.99$, while it is greater than
$10$ when $\gamma=0.75$ and it decreases sensibly with smaller values of $\gamma$.
}

{Figure~\ref{fig3} shows the streamlines and isotherms in the $(r,z)$ plane for the axisymmetric perturbation modes under critical conditions, $\hat{R}=\hat{R}_c$, $\hat{k}=\hat{k}_c$ and $\hat{\Omega}=\hat{\Omega}_c$. For the purpose of drawing the streamlines, we defined a suitable streamfunction, $\psi$, such that
\[
\frac{1}{r}\, \pdv{\psi}{z} \qand - \frac{1}{r}\, \pdv{\psi}{r}
\]
yield the $r$ and $z$ components of the perturbation velocity, respectively.
By decreasing gradually $\gamma$ from $0.99$ to $0.25$, we test of the effect of an increasing curvature of the layer on the shape of the cellular patterns. We see that we have a substantial symmetry of the cells when $\gamma=0.99$ which is slightly broken when $\gamma=0.75$. The asymmetric form of the cells is quite perceivable for $\gamma=0.5$ and becomes even more pronounced for $\gamma=0.25$. There is a tendency for the cells to acquire a boundary layer structure close to the internal boundary at $r=\gamma$ as $\gamma$ decreases. Such a trend is further exploited with very small values of $\gamma$, as illustrated in Fig.~\ref{fig4} where the aspect ratios $\gamma=0.1$ and $\gamma=0.05$ are considered.}

\section{Conclusions}
The stationary and parallel buoyant flow in a vertical porous layer with an annular cross--section has been studied. A uniform internal heat source drives the buoyant flow, leading to instability when its intensity is sufficiently large. The internal and external cylindrical boundaries have been modelled as permeable and with the same given temperature. The governing parameters driving the transition to convective instability are the Rayleigh number, $R$, which is proportional to the heat source intensity, and the aspect ratio, $\gamma$, between the internal radius and the external radius of the annulus. A linear stability analysis has been carried out for the determination of the neutral stability condition and of the critical Rayleigh number for a wide range of aspect ratios, $\gamma$. The main focus has been the evaluation of the effects of the aspect ratio $\gamma$ on the onset of the instability, by considering the limit $\gamma \to 1$ as the reference condition. Indeed, such a limit corresponds to the case of a plane layer which was examined previously by \citet{barletta2017instability}. Thus, gradually decreasing values of $\gamma$ have revealed the effects of an increasing curvature of the layer. Among the most interesting results obtained from this study we mention the following:\\
\textbullet\qquad{}The most unstable perturbation modes are axisymmetric. The distinction between the onset thresholds of axisymmetric and non--axisymmetric perturbation modes tends to be more and more significant as $\gamma$ decreases below unity.\\
\textbullet\qquad{}The evaluation of the critical Rayleigh number revealed that an increasing curvature of the layer generally destabilises the basic buoyant flow. An exception to this trend emerges for very small $\gamma$, namely a parametric domain where the assumed boundary condition at the internal boundary appears to be difficult to implement in a real--world system.

\section*{Acknowledgements}
The authors A. Barletta and B. Pulvirenti acknowledge the financial support from the grant PRIN~2017F7KZWS provided by the Italian Ministry of Education and Scientific Research.

\providecommand{\BIBde}{de~B}

\end{document}